\documentclass[preprint,11pt,aps,nofootinbib,showpacs,tightenlines]{revtex4}
\usepackage{amssymb,latexsym}
\usepackage{amsmath,amsbsy}
\usepackage{epsfig,bm}
\usepackage{graphicx}
\usepackage{comment}
\DeclareMathOperator{\tr}{tr}

\begin{document}
\def\a{{\alpha}}
\def\b{{\beta}}
\def\d{{\delta}}
\def\D{{\Delta}}
\def\e{{\varepsilon}}
\def\g{{\gamma}}
\def\G{{\Gamma}}
\def\k{{\kappa}}
\def\l{{\lambda}}
\def\L{{\Lambda}}
\def\m{{\mu}}
\def\n{{\nu}}
\def\o{{\omega}}
\def\O{{\Omega}}
\def\S{{\Sigma}}
\def\s{{\sigma}}
\def\th{{\theta}}

\def\ol#1{{\overline{#1}}}

\def\Dslash{D\hskip-0.65em /}
\def\Dtslash{\tilde{D} \hskip-0.65em /}

\def\CPT{{$\chi$PT}}
\def\QCPT{{Q$\chi$PT}}
\def\PQCPT{{PQ$\chi$PT}}
\def\tr{\text{tr}}
\def\str{\text{str}}
\def\diag{\text{diag}}
\def\order{{\mathcal O}}

\def\meff{{m^2_{\text{eff}}}}

\def\Meff{{M_{\text{eff}}}}
\def\cF{{\mathcal F}}
\def\cS{{\mathcal S}}
\def\cC{{\mathcal C}}
\def\cE{{\mathcal E}}
\def\cB{{\mathcal B}}
\def\cT{{\mathcal T}}
\def\cQ{{\mathcal Q}}
\def\cL{{\mathcal L}}
\def\cO{{\mathcal O}}
\def\cA{{\mathcal A}}
\def\cV{{\mathcal V}}
\def\cR{{\mathcal R}}
\def\cH{{\mathcal H}}
\def\cW{{\mathcal W}}
\def\cM{{\mathcal M}}
\def\cD{{\mathcal D}}
\def\cN{{\mathcal N}}
\def\cP{{\mathcal P}}
\def\cK{{\mathcal K}}
\def\Qt{{\tilde{Q}}}
\def\Dt{{\tilde{D}}}
\def\psit{{\tilde{\psi}}}
\def\St{{\tilde{\Sigma}}}
\def\cBt{{\tilde{\mathcal{B}}}}
\def\cDt{{\tilde{\mathcal{D}}}}
\def\cTt{{\tilde{\mathcal{T}}}}
\def\cMt{{\tilde{\mathcal{M}}}}
\def\At{{\tilde{A}}}
\def\Qt{{\tilde{Q}}}
\def\cNt{{\tilde{\mathcal{N}}}}
\def\cOt{{\tilde{\mathcal{O}}}}
\def\cPt{{\tilde{\mathcal{P}}}}
\def\cI{{\mathcal{I}}}
\def\cJ{{\mathcal{J}}}

\def\eqref#1{{(\ref{#1})}}

\preprint{UMD-40762-406}
 
\title{Broken Symmetries from Minimally Doubled Fermions}

\author{P.~F.~Bedaque}
\email[]{bedaque@umd.edu}
\affiliation{%
Maryland Center for Fundamental Physics, 
Department of Physics, 
University of Maryland, 
College Park,  
MD 20742-4111, 
USA
}

\author{M.~I.~Buchoff}
\email[]{mbuchoff@umd.edu}
\affiliation{%
Maryland Center for Fundamental Physics, 
Department of Physics, 
University of Maryland, 
College Park,  
MD 20742-4111, 
USA
}

\author{B.~C.~Tiburzi}
\email[]{bctiburz@umd.edu}
\affiliation{%
Maryland Center for Fundamental Physics, 
Department of Physics, 
University of Maryland, 
College Park,  
MD 20742-4111, 
USA
}

\author{A.~Walker-Loud}
\email[]{walkloud@umd.edu}
\affiliation{%
Maryland Center for Fundamental Physics, 
Department of Physics, 
University of Maryland, 
College Park,  
MD 20742-4111, 
USA
}



\begin{abstract}
Novel chirally symmetric fermion actions containing the minimum amount of fermion doubling have been recently proposed in the literature.  We study the symmetries and renormalization of these actions and find that in each case, discrete symmetries, such as parity and time-reversal, are explicitly broken.  Consequently, when the gauge interactions are included, these theories radiatively generate relevant and marginal operators. The restoration of these symmetries and the approach to the continuum limit  thus require the fine-tuning of several parameters.  With some assumptions, we show that this behavior is expected for actions displaying minimal fermion doubling.
\end{abstract}
\maketitle

%
%
\section{Introduction}
A notorious hurdle in the study of lattice field theories is the phenomenon of fermion doubling. In fact, most actions containing only one fermion field per lattice site while retaining at least a remnant chiral symmetry describe several fermion species in the continuum limit. The presence of these extra degrees of freedom is, to a certain degree, unavoidable.  The Nielsen-Ninomiya~\cite{Nielsen:1980rz,Nielsen:1981xu,Nielsen:1981hk} theorem guarantees that under certain very general assumptions, chirally symmetric actions lead to a doubling of the number of fermions one would naively expect.  Fermion actions which bypass the Nielsen-Ninomiya no-go theorem are known but are computationally very expensive.  For actions possessing hypercubic symmetry there is one fermion doubling for every space time direction and thus, in four dimensions, one Dirac fermion on each lattice site leads to sixteen Dirac fermions in the continuum limit. Their use in QCD where only two (or three) light fermions exist is problematic. In the past, Karsten~\cite{Karsten:1981gd} and Wilczek~\cite{Wilczek:1987kw} have suggested four-dimensional actions with the minimal number of doublings allowed by the Nielsen-Ninomiya theorem, namely, one doubling, corresponding to two fermions in the continuum.  Unfortunately, these actions require the fine tuning of some parameters to achieve the continuum limit, which is a daunting numerical task, and have not been extensively used.  The subject was recently revived with the suggestion by Creutz~\cite{Creutz:2007af} of a graphene inspired class of actions, some of them defined on non-hypercubic lattices, possessing both chiral symmetry and the minimal fermion doubling. Soon afterwards Bori\c ci~\cite{Borici:2007kz} suggested a related action and clarified some points of Creutz's proposal.
 
In this letter, we point out that these actions~\cite{Karsten:1981gd,Wilczek:1987kw,Creutz:2007af,Borici:2007kz} break discrete symmetries like parity and time reversal and therefore when gauge interactions are included, one should expect to 
fine tune relevant and marginal operators in order to have a continuum limit.  The need for fine tuning, of course,  hinders the usability of these actions in practical calculations.  Before presenting a systematic analysis, it is useful to make our main point using the Bori\c ci action as an example, which, in momentum space, can be written in the form
\begin{equation}
aD_B(p) = -  i \gamma_4 \frac{4}{\sqrt{2}}
	+i \gamma_4 \sum_{\mu=1}^4 \cos(a p_\mu) 
	+ i \sum_{j=1}^3 \gamma_j s_j(ap),
\end{equation} 
where
\begin{equation}\label{eq:borici1}
\left(\begin{matrix}
s_1(ap)\\
s_2(ap)\\
s_3(ap)\\
\end{matrix}\right) 
=
\left(\begin{matrix}
+1&+1&-1&-1\\
+1&-1&-1&+1\\
+1&-1&+1&-1\\
\end{matrix}\right) 
\left(\begin{matrix}
\sin(ap_1)\\
\sin(ap_2)\\
\sin(ap_3)\\
\sin(ap_4)\\
\end{matrix}\right). 
\end{equation} 
This action breaks the parity symmetry, under which 
$\psi(\vec{p}, p_4)\rightarrow \gamma_4 \psi(-\vec{p}, p_4)$, 
charge conjugation,
$\psi(\vec{p}, p_4)\rightarrow C \bar\psi^T(\vec{p}, p_4)$, 
and  time-reversal, 
$\psi(\vec{p}, p_4)\rightarrow \gamma_5\gamma_4 \psi(\vec{p},- p_4)$. 
Consequently, in the effective (Symanzik) action valid at distances larger than the lattice spacing 
$a$ 
but smaller than the confinement radius, the time-reversal breaking operator
\begin{equation}
 \frac{1}{a}\, \bar\psi\, i  \gamma_4 \, \psi 
\end{equation} 
appears in the theory.   This relevant operator is not part of the continuum limit of QCD and therefore requires its explicit inclusion in the action with a finely tuned coefficient.  This is an example of the fine tuning problem alluded to above. The general analysis of the symmetries and the operator content of the Symanzik action of these theories requires more care and is the central subject of this paper. We will see that these novel actions require multiple fine tunings of dimension three and four operators.

We begin with a general analysis of the Bori\c{c}i and Creutz actions in Sec.~\ref{Action}, 
rewriting these actions to expose their flavor structure.
This allows us to identify symmetries of the action and construct the first few terms of the 
Symanzik effective continuum theory, which is done in Sec.~\ref{Symanzik}.  
In Sec.~\ref{Wilczek}, we comment on the analysis for the Wilczek action.  
We conclude in Sec.~\ref{End} by arguing that any chirally symmetric action of this type 
with minimal doubling will also require the fine tuning of relevant operators.

%
%
\section{The Bori\c{c}i and Creutz Lattice Actions} \label{Action}

Here we consider the fermion actions proposed by Bori\c{c}i~\cite{Borici:2007kz}
and Creutz~\cite{Creutz:2007af}. 
These actions can be written in the general form
\begin{eqnarray} 
S_{BC} 
&=& 
\frac{1}{2}
\sum_{x, \mu}
\Bigg[
\ol \psi (x)
\Big( \Xi_\mu +  i B  \gamma_4 \Big)
\psi (x + \mu)
-
\ol \psi (x)
\Big( \Xi_\mu -  i B \gamma_4 \Big)
\psi (x - \mu )
-
2 B C
\ol \psi (x) 
i \gamma_4 
\, \psi(x)
\Bigg], 
\notag \\
\label{eq:C}
\end{eqnarray}
where the matrices $\Xi_\mu$ are linear combinations 
of the Dirac matrices $\gamma_\mu$, 
\begin{equation}
\Xi_{\mu} = \sum_{j=1}^3 A_{\mu j} \gamma_{j}
,\end{equation}
with
\begin{equation}
A
= 
\begin{pmatrix}
+1 & +1 & +1 \\
+1 & -1 & -1 \\
-1 & -1 & +1 \\
-1 & +1 & -1
\end{pmatrix}
.\end{equation}
The notation for
$\Xi_\mu$ 
is only a shorthand because 
$\Xi_\mu$ 
does not transform as a four-vector under the spinor
representation of the Lorentz group.
The Bori\c{c}i-Creutz action depends on two parameters 
$B$, 
and 
$C$. 
When  
$B = 1$, and 
$C = 1 / \sqrt{2}$,
Eq.~\eqref{eq:C}
reduces to the Bori\c{c}i action.

This action, Eq.~\eqref{eq:C}, can be diagonalized by expanding the field 
$\psi(x)$ in Fourier modes.  
With
\begin{equation}
	\psi(x) = 
		\int_{-\pi}^{\pi} \frac{d^4 p} {(2 \pi)^4}
		e^{i p_\mu x_\mu}
		\tilde{\psi} (p)\, ,
\end{equation}
we find
\begin{equation}
S_{BC} 
=
\int_{-\pi}^{\pi} \frac{d^4 p} {(2 \pi)^4}
\ol \psit (p)
D^{-1}_{BC}(p)
\tilde{\psi} (p)
,\end{equation}
where the propagator 
$D_{BC}(p)$ 
is given by
\begin{equation} 
D_{BC}(p) 
= 
\Bigg[ 
\sum_\mu
\Big(
i \Xi_\mu 
\sin p_\mu
+ 
i B \gamma_4 (\cos p_\mu - C)
\Big)
\Bigg]^{-1}
\label{eq:BCprop}
.\end{equation}
From the last term in the propagator it is clear
that discrete symmetries 
$C$ 
and 
$T$ 
are broken.
Furthermore 
$\Xi_4$ 
is odd under parity, while 
$\sin p_4$ 
is even leading to broken 
$P$ 
in Eq.~\eqref{eq:BCprop}.

Provided 
$B\neq 0$, 
and 
$0 < C < 1$, 
there are exactly two poles of the propagator
corresponding to massless fermions,
namely at
$p^{(1)}_\mu = (\tilde p, \tilde p, \tilde p, \tilde p)$
and
$p^{(2)}_\mu =  - p^{(1)}_\mu$,
where 
$\cos \tilde p = C$.
From the momentum space action, it is straightforward
to define flavored quark fields. Momentum modes near 
$p_\mu^{(1)}$ 
are described by the field
$\tilde{Q}^{(1)}(p)$, 
while those near 
$p_\mu^{(2)}$
are described by 
$\tilde{Q}^{(2)}(p)$. 
The exact partition of the Brillouin zone into these two non-overlapping
regions is irrelevant.  
$\tilde{Q}^{(1)}(p)$ 
should have support in a momentum hypersphere  
centered at 
$p_\mu^{(1)}$,
and 
$\tilde{Q}^{(2)}(p)$ 
should have support in a hypersphere about 
$p_\mu^{(2)}$.
\footnote{%
Notice that a free valence quark of the first variety, for example, can thus be constructed from an operator of the form
\begin{equation}
Q^{(1)}(x) =  
	\sum_{y}
	\int_{\cD} 
	\frac{d^4 p}{(2 \pi)^4}
	e^{i p_\mu (x_\mu - y_\mu)}
	\psi (y)
	\notag
,\end{equation}
where 
$\cD$ 
is a momentum space hypersphere centered about 
$p^{(1)}_\mu$.  
When the theory is gauged, a Wilson line is needed for gauge invariance.  Calculation of correlation functions
of definite flavor requires computations at the level of all-to-all propagators.} %
The union of the two partitions must equal the full Brillouin zone. 
In this scheme, the action can be decomposed into two terms
\begin{equation}
S_{BC} 
= 
\int_{-\pi}^\pi \frac{d^4 p}{(2 \pi)^4}
\Big[
\ol \Qt {}^{(1)} (p)
D^{-1}(p)
\Qt {}^{(1)} (p)
+
\ol \Qt {}^{(2)} (p)
D^{-1}(p)
\Qt {}^{(2)} (p)
\Big]
.\end{equation}
It is convenient to shift both terms
so that the momenta are measured from zero, i.e.
\begin{equation} \label{eq:decomp}
S_{BC} 
= 
\int_{-\pi}^\pi \frac{d^4 q}{(2 \pi)^4}
\Big[
\ol \Qt {}^{(1)} (q + p_{\mu}^{(1)})
D^{-1}(q + p_{\mu}^{(1)})
\Qt {}^{(1)} (q + p_{\mu}^{(1)})
+
\ol \Qt {}^{(2)} (q - p_{\mu}^{(1)})
D^{-1}(q - p_{\mu}^{(1)})
\Qt {}^{(2)} (q -  p_{\mu}^{(1)})
\Big]
,\end{equation}
where periodicity of the momentum fields allows
the integration to remain over a single Brillouin zone. 
The propagator near $\pm p_{\mu}^{(1)}$ has the form
\begin{equation}
D(q \pm p_{\mu}^{(1)}) 
= 
\Bigg[ 
\sum_{\mu}
i  \sin q_\mu \left( C \Xi_\mu \mp BS \gamma_4 \right)
\pm
\sum_{\mu}
i ( \cos q_\mu - 1) \left(  S \Xi_\mu \pm B C \gamma_4 \right)
\Bigg]^{-1}
.\end{equation}
Above $S$ is given by $S = \sqrt{1 - C^2}$.

We can compactly write the action to expose
flavor symmetries with the definition of an
isodoublet $\Qt$, 
\begin{equation}
\Qt (q)
= 
\begin{pmatrix}
\Qt {}^{(1)}(q + p^{(1)}) \\
T \Qt {}^{(2)} (q - p^{(1)}) 
\end{pmatrix}
,\end{equation}
where $T$ is a matrix evoking the similarity transformation:
$\{ \gamma_j \to \gamma_j, \gamma_4 \to -\gamma_4 \}$. 
Thus we find
\begin{eqnarray} \label{eq:Creutzflavour}
S_{BC}
&=&
\int_{-\pi}^\pi \frac{d^4 q}{(2 \pi)^4}
\sum_{\mu}
\Bigg\{
\ol \Qt (q)
\Big[
(C \Xi_\mu - B S \gamma_4 ) \otimes 1
\Big]
i \sin q_\mu \,
\Qt (q) 
\notag \\
&&
\phantom{spacerspacer}+ 
\ol \Qt  (q)
\Big[
(S \Xi_\mu + B C \gamma_4 )  \otimes \tau^3
\Big]
i (\cos q_\mu -1)
\Qt (q)
\Bigg\}.
\end{eqnarray}
The isospin matrix 
$\tau^3$ 
has the familiar definition, 
$\tau^3 = \diag (1, -1)$.

At this point, 
we identify a new set of Dirac matrices 
$\gamma'_\mu$, 
given by
 $\gamma'_\mu = \sum_\nu \gamma_\nu a_{\nu \mu}$, 
with
\begin{equation}
a
= 
\frac{1}{2}
\begin{pmatrix}
+1 & +1 & -1 & -1 \\
+1 & -1 & -1 & +1 \\
+1 & -1 & +1 & -1 \\
-1 & -1 & -1 & -1
\end{pmatrix}
.\end{equation}
Due to the orthogonality of $a_{\mu \nu}$, the $\gamma'_\mu$
is indeed a set of Dirac matrices~\cite{Borici:2007kz}.
We then rewrite the Bori\c{c}i-Creutz action in the form
\begin{eqnarray} \label{eq:Creutzflavor}
S_{BC}
&=&
\int_{-\pi}^\pi \frac{d^4 q}{(2 \pi)^4}
\sum_{\mu}
\Bigg\{
\ol \Qt (q)
\Big[
\left(
2 C \gamma'_\mu - \frac{1}{2} (C - B S) 
\sum_\nu \gamma'_\nu 
\right) \otimes 1
\Big]
i \sin q_\mu \,
\Qt (q) 
\notag \\
&&
\phantom{spacerspacer}+ 
\ol \Qt  (q)
\Big[
\left(
2 S \gamma'_\mu - \frac{1}{2} (S + B C) 
\sum_\nu \gamma'_\nu 
\right) \otimes \tau^3
\Big]
i (\cos q_\mu -1)
\Qt (q)
\Bigg\}.
\end{eqnarray}
One can produce orthogonal axes of the continuum 
action with the choice 
$C = B S$. 
For $B=3$, this corresponds 
to the value $C = 3 / \sqrt{10}$ as shown in~\cite{Creutz:2007af}.
For $B=1$, this corresponds to $C = 1 / \sqrt{2}$, and results 
in the Bori\c{c}i action~\cite{Borici:2007kz}. For an arbitrary 
value of $B$, the choice $C = B / \sqrt{1 + B^2}$ produces
orthogonal axes.
By defining the 
$\gamma'_\mu$ 
matrices, we have absorbed
the momentum rotation into the Dirac algebra. 
After rescaling the fields,
the orthogonal axis
Bori\c{c}i-Creutz action has the form
\begin{eqnarray} \label{eq:Croot10}
S_{BC}^{\text{orthog}}
&=&
\int_{-\pi}^\pi \frac{d^4 q}{(2 \pi)^4}
\sum_{\mu}
\Bigg\{
\ol \Qt (q)
\Big(
\gamma'_\mu  \otimes 1
\Big)
i \sin q_\mu \,
\Qt (q) 
\notag \\
&&
\phantom{spacerspacer}+ 
\frac{1}{B} 
\, \ol \Qt  (q)
\Big[
\left(
\gamma'_\mu - \frac{1}{4} (1+ B^2) 
\sum_\nu \gamma'_\nu 
\right) \otimes \tau^3
\Big]
i (\cos q_\mu -1)
\Qt (q)
\Bigg\}.
\end{eqnarray}
The first term in Eq.~\eqref{eq:Croot10} 
is hypercubic invariant but the second is not. 
Thus no additional symmetry emerges upon
choosing orthogonal axes.

When 
$C \neq BS$, 
the lattice hyperplanes 
do not constitute an orthogonal coordinate 
system for the particle momenta. 
In the continuum limit, the action
Eq.~\eqref{eq:Creutzflavor}
tends to 
$i \gamma_\mu k_\mu$,
where the 
$k_\mu$ 
specify the particle momenta
in an orthogonal system. 
This yields $k_\mu = R_{\mu \nu} q_\nu$, 
where as a matrix, $R^{-1}$ is given by
\begin{equation}
R^{-1}
=
\frac{1}{4 C}
\begin{pmatrix}
+1 & +1 & +1 &  - \frac{C}{BS} \\
+1 & -1 & -1 &  - \frac{C}{BS} \\
-1 & -1 & +1 &  - \frac{C}{BS} \\
-1 & +1 & -1 &  - \frac{C}{BS} \\
\end{pmatrix}
.\end{equation}
In terms of physical particle momenta, we have
\begin{eqnarray} \label{eq:Creutzflavor2}
S_{BC}
&=&
\int  \frac{d^4 k}{(2 \pi)^4}
\sum_{\mu}
\Bigg\{
\ol \Qt (k)
\Big[
\left(
2 C \gamma'_\mu - \frac{1}{2} (C - B S) 
\sum_\nu \gamma'_\nu 
\right) \otimes 1
\Big]
i \sin ( R^{-1} k)_\mu \,
\Qt (k) 
\notag \\
&&
\phantom{spacerspa}+ 
\ol \Qt  (k)
\Big[
\left(
2 S \gamma'_\mu - \frac{1}{2} (S + B C) 
\sum_\nu \gamma'_\nu 
\right) \otimes \tau^3
\Big]
i \Big(\cos ( R^{-1} k)_\mu -1 \Big)
\Qt (k)
\Bigg\}.
\end{eqnarray}
The symmetries of the properly gauged version of 
Eq.~\eqref{eq:Creutzflavor2}
are specifically as follows:
\begin{enumerate}

\item
Gauge invariance

\item
Discrete spacetime translation invariance

\item
$U(1)_B$ baryon number

\item
$U(1)_L \otimes U(1)_R$ 
chiral symmetry corresponding to 
$\tau^3$

\item
Combined 
$CPT$ 
invariance (but none individually)

\item
Combined 
$IC$ 
invariance, where $I$ is the isospin rotation 
$I = \exp \left( \pm i  \pi \tau^1 / 2 \right)$

\item
$S_4$ invariance under permutations involving all four hyperplane axes.
 
\end{enumerate}
The last symmetry can be translated into 
fairly complicated transformations among 
the physical axes of particle motion.
We have, however, found it easier to work with the 
original hyperplane axes.  
While there is an axial 
$U(1)_A$ 
symmetry possessed by 
Eq.~\eqref{eq:Creutzflavor}, near the continuum it will 
be spoiled by the anomaly.
Lastly spontaneous symmetry breaking reduces  the 
chiral symmetry down to 
$U(1)_V$ 
flavor symmetry generated by 
$\tau^3$.
It might be possible that additional non-standard symmetries
emerge for particular values of 
$B$ 
and 
$C$, 
such as for the values 
$B = \sqrt{5} \cot (\pi / 5)$ 
and
$C = \cos(\pi / 5)$
suggested by analogy with graphene~\cite{Creutz:2007af}.

%
%
\section{Symanzik Effective Theory} \label{Symanzik}

To investigate discretization errors, 
we consider the continuum limit.
Using the seven symmetries of the Bori\c{c}i-Creutz action, 
it is straightforward to construct the Symanzik effective theory, 
valid at distances larger than the lattice spacing but smaller than the confinement radius%
~\cite{Symanzik:1983dc,Symanzik:1983gh}. 
The general form of the Symanzik theory is
\begin{equation}
\cL_{\text{eff}} 
=
\sum_{n = 3}^\infty 
\sum_{j}
a^{n-4}
c_{n}^{(j)} \cO_{n}^{(j)}
,\end{equation}
where the sum on $n$ is over the operator dimension
and the sum on $j$ is over all operators of a given dimension.
As the symmetries of the Bori\c{c}i-Creutz action are non-standard, 
however,
we shall begin first with a na\"ive demonstration to produce a term generated from 
radiative corrections.

Consider the lattice spacing and gauge coupling sufficiently small.
Accordingly we can treat the theory as weakly coupled. 
Moreover, 
we can expand the action in Eq.~\eqref{eq:Creutzflavor2}
in powers of the lattice spacing and perform a Fourier
transform to position space. 
Up to 
$\cO(a)$, 
the theory becomes (in dimensionful units)
\begin{equation} \label{eq:Leff}
\cL_{\text{eff}} 
=
\ol Q (\gamma_\mu \otimes 1) D_\mu Q
- 
\frac{1}{4} 
F_{\mu \nu} F_{\mu \nu}
- a \frac{i}{2} 
\, \ol Q 
\left(  2 S \gamma'_\mu - \frac{1}{2} (S + BC) \sum_\nu \gamma'_\nu \right) 
\otimes \tau^3  
(R^{-1} D)_\mu (R^{-1}D)_\mu Q
,\end{equation}
with the gauge covariant derivative given by 
$D_\mu = \partial_\mu - i g A_\mu$, where $A_\mu$
is valued in the algebra of the gauge group.
The gauge field strength tensor is given through 
the relation $[ D_\mu, D_\nu] = - i g F_{\mu \nu}$.

\begin{figure}[!ht]
\epsfig{file=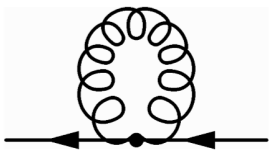,width=3.5cm}
\caption{
Self-energy tadpole which generates 
contributions to a relevant operator. 
}
\label{f:tadpole}
\end{figure}

Now we use the effective action in Eq.~\eqref{eq:Leff}
to  investigate the fermion self-energy. 
Working perturbatively
in the coupling constant, there are discretization corrections
which arise from an insertion of the $\cO(a)$ dimension five operator
$\cO_5 
=  
\ol Q ( i \zeta_\mu \otimes \tau^3 ) (R^{-1} D)_\mu (R^{-1} D)_\mu Q$.
Here $\zeta_\mu$ is just a replacement for the combination of
$\gamma'_\mu$ matrices appearing in Eq.~\eqref{eq:Leff}.
The $\zeta_\mu$ do not in general satisfy the Dirac algebra.
A self-energy diagram generated from inserting this operator
is shown in Figure~\ref{f:tadpole}.
This tadpole is straightforward to evaluate. Schematically we find
\begin{equation}
 \ol u(p)  \int \frac{d^4 q}{q^2} 
\sum_{\mu} a \, \alpha_s  (i \zeta_\mu \otimes \tau^3) 
u(p)
\sim
\frac{\alpha_s}{a} \,
\ol u(p) \sum_\mu
( i \zeta_\mu \otimes \tau^3 )
u(p)
,\end{equation}
which corresponds to a relevant operator in the Symanzik action
of the form
\begin{equation}
\mathcal{O}_3 
=
\sum_\mu
\ol Q 
(i \zeta_\mu \otimes \tau^3 )
Q
.\end{equation}
Now it could be possible that additional 
$\cO(\alpha_s)$ 
diagrams happen to cancel this 
$\cO(a^{-1})$ 
term.%
\footnote{%
There are also operators of higher dimension that contribute to the fermion self-energy at 
$\cO(a^{-1})$. The reason being that higher dimensional operators can produce terms $\sim (a q)^n$, where $q$ is a loop momentum that can reach the cutoff.  Thus all such terms are order unity.
One must use lattice perturbation theory to determine all contributions.  Our example is meant to demonstrate the existence of such terms, not to determine their coefficients rigorously.}
However, without a symmetry protecting this term from being generated by the interactions, it likely receives contributions from all orders in the gauge coupling and appears in the Symanzik action with a non-perturbatively determined coefficient.  One can verify this operator, $\cO_3$, satisfies the seven symmetries of the Bori\c{c}i-Creutz lattice action.

Using  the seven symmetries of the Bori\c{c}i-Creutz action, 
we construct all relevant, $\cO_3^{(j)}$, and marginal, $\cO_4^{(j)}$, operators
present in the Symanzik effective theory $\cL_\text{eff}$, 
\begin{equation}
\cL_\text{eff}
= 
\frac{1}{a}
\sum_{j} c_3^{(j)} \cO_3^{(j)}
+ 
\sum_j c_4^{(j)} \cO_4^{(j)}
+ 
\cO(a)
.\end{equation}
The relevant operators are
\begin{eqnarray*}
\cO_3^{(1)} 
&=&  
\ol Q (i \gamma_4 \otimes \tau^3 ) Q
\\
\cO_3^{(2)} 
&=&  
\ol Q ( \gamma_4 \gamma_5 \otimes \tau^3 ) Q
.\end{eqnarray*}
One notes that 
$\sum_\nu \gamma'_\nu =  - 2 \gamma_4$,
which makes obvious the permutation symmetry
of the original hyperplane axes.
The operator $\cO_3^{(1)}$ can 
be eliminated by a field redefinition,
see Sect.~\ref{Wilczek} for a discussion
of the analogous operator for the Wilczek action.

\begin{figure}[!t]
\epsfig{file=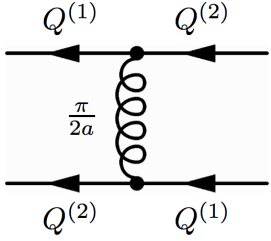,width=3.5cm}
\caption{
High momentum gluon interaction
which leads to flavor exchange.
Here the gluon momentum is 
$\frac{\pi}{2 a}$ which is specific
to the Bori\c i action. In general, 
the flavor changing momentum
is $\frac{2}{a} \cos^{-1} C$.
}
\label{f:OGE}
\end{figure}

The marginal operators in the Symanzik theory are
\begin{eqnarray*}
\cO_4^{(1)}
&=&
\ol Q ( \gamma_\mu \otimes 1) D_\mu  Q 
\\
\cO_4^{(2)}
&=&
\ol Q ( \gamma_4 \otimes 1) D_4  Q 
\\
\cO_4^{(3)}
&=&
\ol Q (i \gamma_\mu \gamma_5 \otimes 1) D_\mu Q
\\
\cO_4^{(4)}
&=&
\ol Q (i \gamma_4 \gamma_5 \otimes 1) D_4 Q
\\
\cO_4^{(7)}
&=& 
F_{\mu \nu} F_{\mu \nu}
\\
\cO_4^{(8)}
&=& 
F_{\mu \nu} \tilde{F}_{\mu \nu}
\\
\cO_4^{(9)}
&=& 
F_{4 \mu} F_{4 \mu}
\\
\cO_4^{(10)}
&=& 
F_{4 \mu} \tilde{F}_{4 \mu}
,\end{eqnarray*}
where
$\tilde{F}$ 
is the dual tensor to 
$F$, 
namely
$\tilde{F}_{\mu \nu} = \frac{1}{2} \varepsilon_{\mu \nu \rho \sigma} F_{\rho \sigma}$. 
Beyond this order there is an explosion in the number of 
operators. For example, at 
$\cO(a)$ 
there are irrelevant operators that lead to chromoelectric dipole moments.
At $\cO(a^2)$, one has various four quark operators. 
Analogous to the case of taste breaking in staggered fermions, 
there are flavor changing interactions. 
We exemplify this in Figure~\ref{f:OGE}, where a 
high momentum gluon exchange leads to mixed-flavor
four-quark interactions. Notice that the individual flavor
numbers are conserved in accordance with the 
$U(1)_B \otimes U(1)_V$
symmetry of the action.
%
%
%
%
%

%
%
\section{The Wilczek Action} \label{Wilczek}

Karsten~\cite{Karsten:1981gd} and Wilczek~\cite{Wilczek:1987kw} pointed out, over twenty years ago, the possibility of minimal 
doubling with an explicit construction.
The Wilczek action has the form
\begin{eqnarray}
S_W
&=& 
\frac{1}{2}
\sum_{x,\mu}
\left[
\ol \psi (x) 
\gamma_\mu
\psi (x+\mu)
- 
\ol \psi (x)
\gamma_\mu
\psi (x-\mu)
\right]
+
3 \lambda
\sum_{x}
\ol \psi (x)
i \gamma_4 
\psi(x)
\notag \\
&& \phantom{space}
-
\frac{\lambda}{2}
\sum_{x,j}
\left[
\ol \psi(x) i \gamma_4 \psi (x + j)
+ 
\ol \psi(x) i \gamma_4 \psi (x - j)
\right]
,\end{eqnarray}
with 
$\lambda >\frac{1}{2}$
as a parameter.
The action is diagonal in momentum space
\begin{equation}
S_W 
= 
\int_{-\pi}^{\pi} \frac{d^4 k}{(2 \pi)^4}
\ol \psit (k) 
D^{-1}(k)
\psi(k)
,\end{equation}
with the propagator 
$D(k)$ 
given by
\begin{equation}
D(k) 
=
\left[
\sum_\mu i \gamma_\mu \sin k_\mu
- 
i \gamma_4 \lambda
\sum_{j}
( \cos k_j - 1)
\right]^{-1}
.\end{equation}
There are exactly two poles of the propagator, 
namely at the values
$k_\mu^{(1)} = (0,0,0,0)$,
and 
$k_\mu^{(2)} = (0,0,0,\pi)$.
As this action has been analyzed 
in detail previously~\cite{Pernici:1994yj},
we shall not pursue the decomposition
of the fields in terms of flavor and chirality.
Instead we will focus on the discrete
symmetries of the action:
\begin{enumerate}
\item
Cubic invariance

\item $CT$ invariance

\item $P$ invariance

\item $\Theta$ invariance,

\end{enumerate}
where 
$\Theta$ 
is the time link-reflection symmetry operator. 
For the Bori\c{c}i and Creutz actions, 
$\Theta$ is not a symmetry.

Given these discrete symmetries, it is straightforward to write
down the relevant and marginal terms in the Symanzik action. 
There is one relevant operator
\begin{equation}
\cO_3 
= 
\ol \psi 
i \gamma_4
\psi
\notag
,\end{equation}
and several marginal operators
\begin{align*}
\cO_4^{(1)} &=  F_{\mu\nu} F_{\mu\nu} \\
\cO_4^{(2)} &=  F_{4\mu} F_{4\mu} \\
\cO_4^{(3)} &= 
\bar{\psi}\, D_4 \gamma_4\, \psi \\
\cO_4^{(4)} &=  \bar \psi D_\mu \gamma_\mu  \psi
\end{align*}
As with anisotropic lattices, the breaking of hypercubic symmetry down to
cubic symmetry requires a speed of light tuning.
It has been argued in Ref.~\cite{Pernici:1994yj} that the relevant operator can be absorbed by a field redefinition, 
namely
\begin{equation} \label{eq:fieldredef}
\psi 
\longrightarrow
\exp \left(- \frac{i c_3 x_4}{a ( 1 + c_4^{(3)})} \right) \psi
,\end{equation}
where 
$c_3$
is the non-perturbatively determined 
coefficient of the operator 
$\cO_3$ 
in the Symanzik action, 
while $c_4^{(3)}$ that of $\cO_4^{(3)}$. 
The field redefinition, however, modifies the 
boundary condition and just moves the fine tuning problem from the bulk of the action 
to the boundary. 
The physical reality of this term can be ascertained by noticing that it plays the role of  
an imaginary chemical potential.
The relevant operator thus cannot 
be avoided in the case of the Wilczek action.

\section{Conclusion} \label{End}

We have investigated above different
lattice actions with minimal fermion doubling.
In particular, we demonstrated that the 
Bori\c{c}i-Creutz action has an exact
$U(1)_L \otimes U(1)_R$ 
chiral symmetry. 
For each action, however, there are relevant and marginal 
operators appearing in the Symanzik effective theory.
Breaking of discrete symmetries, such as parity 
and time-reversal, is the culprit for the appearance of such operators.
The Wilczek action has the least number of such
operators.
\begin{figure}[!t]
\epsfig{file=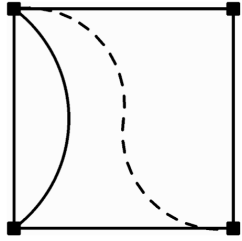,width=3.1cm}
\caption{
Two dimensional slice of momentum space.
Solid squares denote poles, or potential poles of the propagator.
}
\label{f:poles}
\end{figure}

A natural question arising from our analysis is whether a chirally symmetric action with minimal fermion doubling which does not generate dimension three operators is possible. This would indeed occur if such a putative action preserved $PT$ symmetry as that symmetry is sufficient to forbid the dimension three operators listed. It is, however, easy to see that such an action does not exist.
To have minimal fermion doubling, 
it is clear that 
hypercubic invariance must be broken.
Heuristically one can argue that parity or time-reversal 
must also be broken from Figure~\ref{f:poles}.
Here we plot a two-dimensional slice of momentum space.
A shift in momentum space can always be made so that
one pole of the propagator is at a corner of the slice depicted.
Poles of the propagator are shown in the case of ordinary fermion
doubling,
and two minimal doubling scenarios.
Notice that in the absence of symmetric doublers, 
there is always an asymmetry about one axis.
Hence, depending on which axis is asymmetric, 
either 
$P$, 
or 
$T$
is broken. 
The dashed line shows a hypothetical scenario in which the product 
$PT$
is preserved. 
If we write the general chirally symmetric 
momentum space action as
\begin{equation}
S = 
\sum_\mu  i \gamma_\mu f_\mu(p_\nu)
+
\sum_\mu  \gamma_\mu \gamma_5 g_\mu(p_\nu)
,\end{equation}
then we require that each 
$f_\mu$ and $g_\mu$
be odd, e.g..~%
$f_\mu (- p_\nu) = - f_\mu( p_\nu)$,
so that 
$PT$
is preserved. 
Furthermore the action must  be local
which requires the continuity of 
$f_\mu$, $g_\mu$
and periodicity.
The only odd functions which are continuous
and periodic within the Brillouin zone vanish
at the boundaries $\pi/a$ and $-\pi/a$ (as well as at the origin). 
Thus there is no $PT$ symmetric action which
has only minimal doubling.
We have reasoned
that
chirally symmetric
minimal fermion doubling requires
broken 
$P$, 
or 
$T$, 
and broken 
$PT$. 
Without these symmetries intact, 
however, 
dimension three operators are allowed, 
e.g.~%
$\ol \psi i \gamma_j \psi$ for 
$j$ 
spatial, or
$\ol \psi  \gamma_4 \gamma_5 \psi$
in the case of broken
$P$,
and
$\ol \psi  \gamma_j \gamma_5 \psi$
for 
$j$
spatial, or
$\ol \psi i \gamma_4 \psi$
for the case of broken $T$. 
Thus 
chirally symmetric minimal doubling actions, of the form considered here,
require the fine tuning of relevant operators.
There may be additional non-standard 
symmetries, however, that emerge for 
particular values of parameters.
Such symmetries are very interesting 
because they could potentially eliminate 
relevant operators. 
It is thus worthwhile to expose such symmetries.

\begin{acknowledgments}
We thank 
A.~Bori\c{c}i,
M.~Creutz, 
M.~Golterman, 
and 
Y.~Shamir 
for comments on the paper.
This work is supported in part by the 
U.S.~Dept.~of Energy,
Grant No.~DE-FG02-93ER-40762.
\end{acknowledgments}

\appendix

\end{document}